%% file: main.tex
\documentclass[conference]{IEEEtran}
\renewcommand{\baselinestretch}{0.99}
\IEEEoverridecommandlockouts
\usepackage{myStyle}

\AddToShipoutPictureBG*{%
	\AtPageUpperLeft{%
		\setlength\unitlength{1in}%
		\hspace*{\dimexpr0.5\paperwidth\relax}
		\makebox(0,-0.75)[c]{\begin{tabular}{c c}
				Max van Haren, Gaussian Process Position-Dependent Feedforward: With Application to a Wire Bonder, \\
					In {\em IEEE 17th International Conference on Advanced Motion Control}, Padova, Italy, 2022%
			\end{tabular}}
}}

\begin{document}

\title{Gaussian Process Position-Dependent Feedforward: With Application to a Wire Bonder
\thanks{This work was carried out when the first author was an MSc. student, in collaboration with ASM PT. Currently, the first author is a PhD researcher with the IMOCO4.E consortium, which has received funding from the ECSEL Joint Undertaking under grant agreement 101007311 (IMOCO4.E). The Joint Undertaking receives support from the European Union’s Horizon 2020 research and innovation programme. In addition, this work is part of the research programme VIDI with project number 15698, which is (partly) financed by the Netherlands Organisation for Scientific Research (NWO).}}

\author{\IEEEauthorblockN{Max van Haren\IEEEauthorrefmark{1}, Maurice Poot\IEEEauthorrefmark{1}, Dragan Kosti\'c\IEEEauthorrefmark{2}, Robin van Es\IEEEauthorrefmark{2}, Jim Portegies\IEEEauthorrefmark{3} and Tom Oomen\IEEEauthorrefmark{1}\IEEEauthorrefmark{4}}
	\IEEEauthorblockA{\IEEEauthorrefmark{1}Control Systems Technology, Dept. of Mechanical Engineering, Eindhoven University of Technology, \\
	Eindhoven, The Netherlands, email: \texttt{m.j.v.haren@tue.nl}}
	\IEEEauthorblockA{\IEEEauthorrefmark{2}ASM Pacific Technology, Centre of Competency, Beuningen, The Netherlands}		
	\IEEEauthorblockA{\IEEEauthorrefmark{3}CASA, Dept. of Mathematics and Computer Science, Eindhoven University of Technology, Eindhoven, The Netherlands}
	\IEEEauthorblockA{\IEEEauthorrefmark{4}Delft Center for Systems and Control, Delft University of Technology, Delft, The Netherlands}}

\maketitle

\begin{abstract}
Mechatronic systems have increasingly stringent performance requirements for motion control, leading to a situation where many factors, such as position-dependency, cannot be neglected in feedforward control. The aim of this paper is to compensate for position-dependent effects by modeling feedforward parameters as a function of position. A framework to model and identify feedforward parameters as a continuous function of position is developed by combining Gaussian processes and feedforward parameter learning techniques. The framework results in a fully data-driven approach, which can be readily implemented for industrial control applications. The framework is experimentally validated and shows a significant performance increase on a commercial wire bonder.
\end{abstract}

\input{Introduction}
\input{ProblemDefinition}
\input{Approach}

\input{ExperimentalStudy}

\section{Conclusion}
\label{sec:conclusion}
The developed framework models feedforward parameters as a function of position with GPs, resulting in constant performance for any arbitrary position. The training data for the GPs are determined using ILCBF, but can be extended to any feedforward parameter tuning approach. Since the position-dependent effects of motion systems are typically unknown or hard to model, non-parametric regression techniques, such as GPs, work especially well. GPs, in combination with a data-driven feedforward parameter tuning approach, results in a framework where data outweighs assumptions on the position-dependency of a motion system. Lastly, experiments on a complex motion system show that the framework can improve control performance significantly by compensating for, e.g., an unknown position-dependent magnetic flux density. Ongoing work on this topic is directed at optimal and automatic computation of the training positions using sensor placement techniques such as mutual information optimization. 
\section*{Acknowledgment}
The authors wish to thank Kelvin Kai Wa Yan for his support during the experimental study.

\bibliographystyle{IEEEtran}
\bibliography{IEEEabrv,references}

\end{document}

%% file: Introduction.tex
\section{Introduction}
\label{sec:intro}
Increasing performance requirements for motion control leads to situations where position-dependency of mechatronic systems cannot be neglected anymore. An example is the wafer stage, where a flexible mode of a wafer is observed differently for each position \cite{Voorhoeve2021}. Furthermore, H-drive machines, such as large-format printing systems, suffer from position-dependent dynamics due to a changing configuration \cite{Zundert2016}. Traditionally, position-dependency is neglected and the feedforward controller is kept constant over the machine operating range, resulting in suboptimal performance. Recently, due to the developments in computational power, data-driven techniques, such as Gaussian Processes (GPs), are getting more relevant for motion control. \par

Traditional Linear Time-Invariant (LTI) feedforward design attempts to compensate for a known reference signal of a system. Typically, feedforward controllers are based on models. For low frequencies, motion systems can generally be modeled as a rigid-body, resulting in the well-known acceleration feedforward. Acceleration feedforward controllers can directly be extended to compensate for higher-order or non-linear dynamics such as flexible mechanics \cite{Boerlage2004}. The parameters used in traditional LTI controllers can be tuned manually in a straightforward manner \cite{Oomen2020}. Nevertheless, traditional LTI feedforward does not compensate for position-dependent effects. \par

Position-dependent feedforward design can compensate for position-dependent behavior of systems. For this purpose, feedforward parameters can be determined in a grid and estimated with parametric regression techniques such as linear interpolation. However, interpolations have approximation errors since the dependency between position and feedforward parameters is generally unknown. \par

High motion control performance for systems with position-dependent dynamics can furthermore be achieved through the use of Linear Parameter Varying (LPV) or non-linear inversion-based control of the system. First, LTI dynamics can be scheduled according to the current configuration of the LPV system, resulting in high control performance for e.g. wafer stages \cite{Wassink2005} or xy-positioning tables \cite{Toth2011}. Second, data-driven learning techniques, such as Iterative Learning Control (ILC) or adaptive control, can be extended to LPV systems, which results in high performance through learning \cite{deRozario2017,Middleton1988}. Third, a non-linear inversion-based output tracking solution can be applied, resulting in asymptotically exact output tracking \cite{Devasia1996}. LPV model-free approaches are investigated in \cite{Formentin2016}, directly learning LPV controllers from data, but are at present not competitive with model-based designs. The high performance achieved by using LPV or non-linear control typically requires accurate and extensive modeling, that is often very challenging and the high cost and complexity are usually not justified for industrial control applications. \par

Although feedforward design has improved significantly compared with traditional acceleration feedforward, a feedforward with systematic tuning for position-dependent effects, capable of estimation at any arbitrary position, is currently lacking. This paper models feedforward parameters as a continuous function of position through a GP \cite{Rasmussen2004,Pillonetto2014}, which allows for the compensation of position-dependent effects without a full LPV or non-linear model. In addition, a GP is non-parametric and therefore does not require an assumption on the parametr ic form between the position and the feedforward parameters. Gaussian processes have been applied previously in system identification \cite{Chen2012a}, but are not broadly applied to feedforward and are not yet developed for position-dependent systems. In this paper, the feedforward parameters of a system are learned in a trial-to-trial fashion using ILC with Basis Functions (ILCBF) \cite{vandeWijdeven2010}. The contributions include:
\begin{enumerate}
	\item[C1] a generic framework to model feedforward parameters as a function of position using GPs, which can be readily implemented for industrial applications,
	\item[C2] ILCBF to automatically learn feedforward parameters for multiple fixed positions, suitable for industrial machines, that are directly used in the GPs,
	\item[C3] application and validation of the framework to a state-of-the-art industrial experimental setup, showing the capabilities of the framework.
\end{enumerate}
The outline of this paper is as follows. In Section~\ref{sec:problem}, the problem that is considered in this paper is defined. In Section~\ref{sec:approach}, the method for modeling feedforward parameters as a GP and an approach to automatically learn feedforward parameters with ILCBF is described, leading to contributions C1 and C2. In Section~\ref{sec:experiments}, a case study on an experimental setup is performed, constituting contribution C3. Finally, in Section~\ref{sec:conclusion}, concluding remarks are given.

\textbf{Notation:} Systems can be single-input single-output or multiple-input multiple-output with $n_i$ inputs and $n_o$ outputs. All systems are discrete-time, unless stated otherwise, with discrete-time $\dt \in \{0,1,\ldots,N-1\}$. Continuous time systems are transformed in their discrete-time counterpart using finite difference approximation. The trial number is indicated with the index $j$. Signals are assumed to be of length $N$. The weighted 2-norm of a vector $x \in \mathbb{R}^N$ is denoted as $\| x \|_W := \sqrt(x^\top W x)$, where $W \in \mathbb{R}^{N\times N}$ is a weighting matrix. Matrix $A\in\mathbb{R}^{N\times N}$ is positive (semi-)definite if and only if $x^\top A x \geq 0, \; \forall x \neq 0 \in \mathbb{R}^N$ and is denoted as $A\succeq 0$.

%% file: ProblemDefinition.tex
\section{Problem Definition}
\label{sec:problem}
In this section, the problem for determining a position-dependent feedforward controller is formulated. First, a problem setup is given, including the systems considered and the parametrization of the feedforward signal. Second, the three categories of position-dependent effects in mechatronic systems considered are elaborated upon. Finally, a hypothesis of the largest contribution to the position-dependency is made and the problem addressed in this paper is defined.
\subsection{Problem Setup}
\begin{figure}[tbp]
	\centering
	\includegraphics[width=0.8\columnwidth]{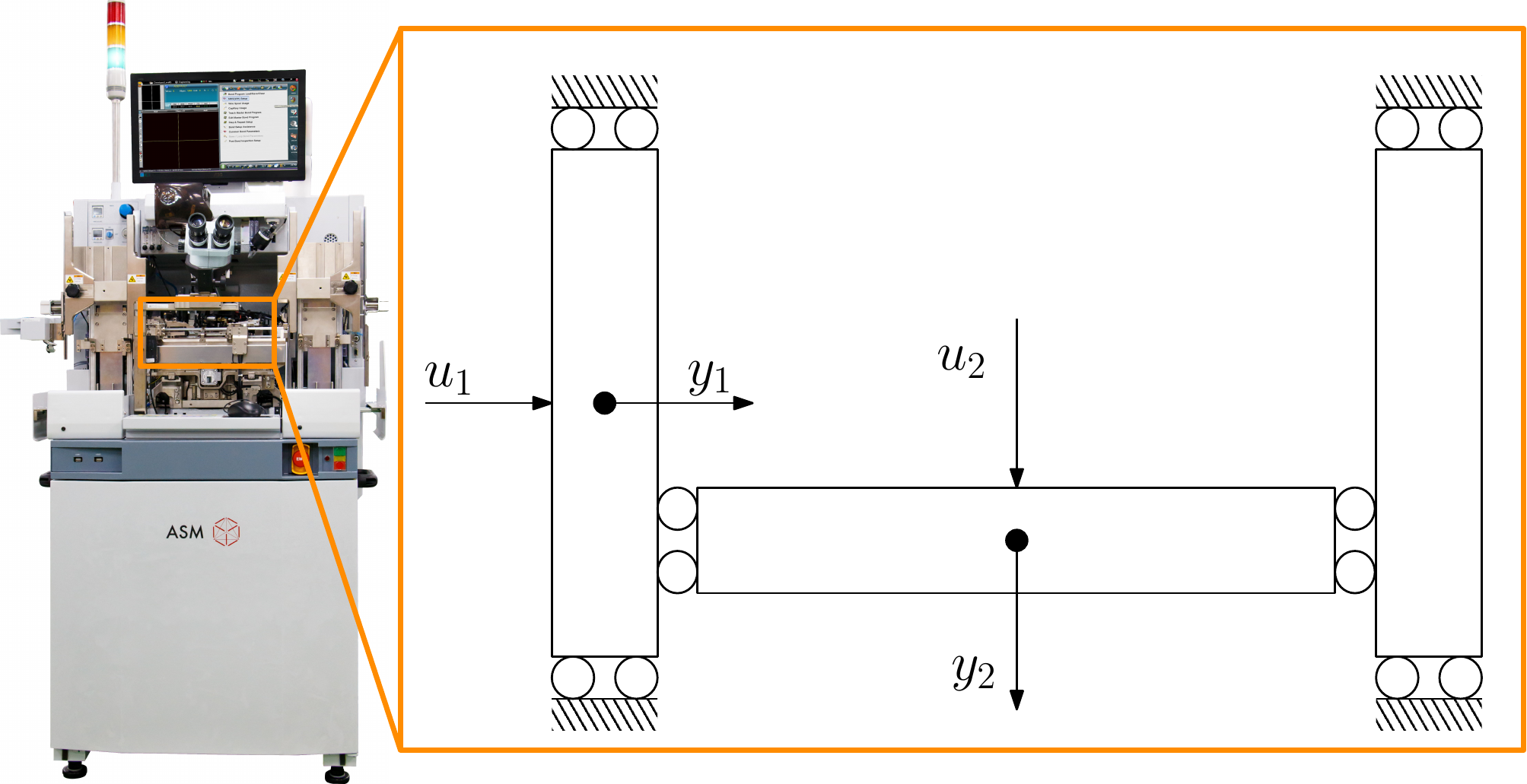}
	\caption{Commercial wire bonder used for the experimental case study. The system might be position-dependent, due to the changing mass distribution.}
	\label{fig:AB383}
\end{figure}

The considered class of position-dependent systems are spatially distributed LTI systems \cite{Moheimani2003}
\begin{equation}
	\label{eq:system}
	y(\dt) = G(\rho,q^{-1})u(\dt),
\end{equation}
with output $y(\dt) \in \mathbb{R}^{N\times n_o}$, input $u(\dt) \in \mathbb{R}^{N\times n_i}$, system $G(\rho,q^{-1})\in\mathbb{R}^{n_o\times n_i}$, the initial position $\rho\in\mathbb{R}^{n_o}$ and $q$ denotes the forward-shift operator, i.e. $q^{-\tau}a(\dt)=a(\dt-\tau)$. Spatially distributed LTI systems are generally applicable to systems with slowly varying position-dependency, i.e., the position-dependency is mostly exerted due to the initial position and not due to the reference signal. The wire bonder in Fig.~\ref{fig:AB383} is a benchmark example, since references are relatively short compared with the machine operating range. \par
\begin{figure}[tbp]
	\centering
	\includegraphics[width = \FigWidth]{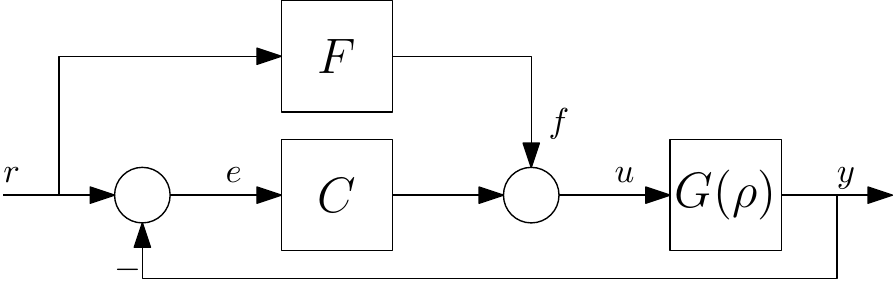}
	\caption{Control structure considered in this paper, with feedforward controller $F$, feedback controller $C$ and the spatially distributed LTI system $G(\rho)$.}
	\label{fig:ControlStructure}
\end{figure}

The applied control structure can be seen in Fig.~\ref{fig:ControlStructure}.
Typically, the design of feedforward controller $F$ tries to reduce the reference-induced error to zero, i.e. $e(\dt) = 0 =r(\dt)-G(\rho,q^{-1})u(\dt)$, resulting in the ideal feedforward controller $F=G^{-1}(\rho,q^{-1})$. Often, the feedforward design is based on the physical model of a system. Consider, e.g., a finite-difference approximation of a mass-damper system
\begin{equation*}
	G_{md}(q^{-1}) = \frac{1}{m \frac{(1-q^{-1})^2}{T_s^2}+d \frac{1-q^{-1}}{T_s}},
\end{equation*}
with $T_s$ the sampling time of the system, $m$ the mass of the system and a damping coefficient $d$. Zero reference-induced error is achieved by the ideal feedforward controller $F(q^{-1}) = m\frac{(1-q^{-1})^2}{T_s^2}+d\frac{1-q^{-1}}{T_s}$. However, due to model uncertainties, the exact system parameters are unknown and therefore the controller takes the form
\begin{equation*}
	F(q^{-1}) = \begin{bmatrix}
		\frac{(1-q^{-1})^2}{T_s^2} & \frac{1-q^{-1}}{T_s}
	\end{bmatrix}\begin{bmatrix}
		\hat{m} \\
		\hat{d}
	\end{bmatrix}= \Psi(q^{-1})\theta,
\end{equation*}
with $\hat{\cdot}$ an estimate or modeled value in the feedforward parameters $\theta \in \mathbb{R}^{n_\theta\times n_i}$ and the basis function matrix $\Psi \in \mathbb{R}^{n_o\times n_\theta}$. Due to the position-dependency of system $G(\rho,q^{-1})$, best performance is achieved by modeling the feedforward parameters $\theta$ as a function of position.
\subsection{Considered Position-Dependent Effects}
Position-dependency of $G(\rho,q^{-1})$ can be caused by several factors, here separated in actuation, mechanical or sensing. First, actuation can cause position-dependency due to for instance cogging \cite{Berkel2007} or varying magnetic flux density \cite{Gieras2012}. Second, mechanical position-dependency can directly affect the feedforward parameters that achieve optimal performance. Examples of such are raster scanning of atomic force microscopy \cite{Butterworth2011} or position-dependent gantry systems, such as the large-format printer \cite{Zundert2016}. Lastly, since sensors are typically attached to the fixed world and motion systems are moving by definition, deformations or movements of the system are observed differently for particular positions. This is for example seen in lithography, where a flexible mode of a wafer is observed differently \cite{Voorhoeve2021}. Due to the high costs associated with modeling each position-dependent effect and manual tuning for all operating positions is infeasible, a data-driven approach for position-dependent feedforward is necessitated.
\subsection{Hypotheses and Problem Definition}
Often, flexible mechanics and the observation thereof are the largest contribution of position-dependency of mechatronic systems, see for instance \cite{Voorhoeve2021,Kontaras2016}. This is anticipated due to the recent developments in lightweight design of mechatronic systems, typically resulting in more flexible structures. Furthermore, actuation can have significant contribution on the position-dependency as well, but is generally compensated for in an earlier step, e.g., by using calibration \cite{Mu2009}. \par
Position-dependent dynamics of system $G(\rho,q^{-1})$ is not compensated for by using constant feedforward parameters. Hence, the aim of this paper is to model feedforward parameters $\theta$ as a continuous function of position, such that it can compensate for any unknown position-dependent effect, without an LPV model or specification of a parametric form, that might lead to estimation errors. 

%% file: Approach.tex
\section{Approach}
\label{sec:approach}
\begin{figure}[!t]
	\centering
	\includegraphics[width=\linewidth]{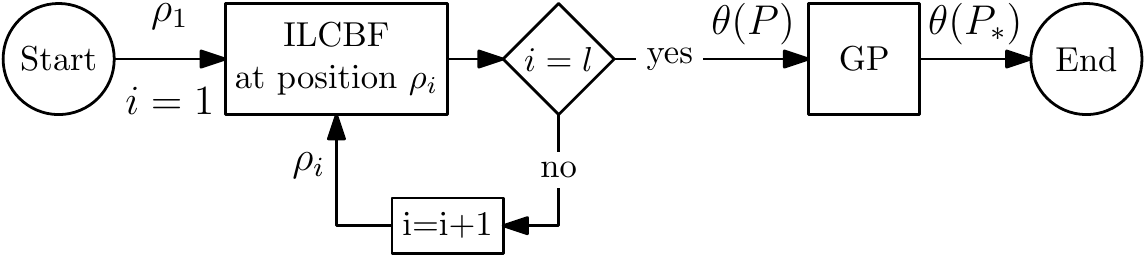}
	\caption{Illustration of the developed framework, where feedforward parameters are being modeled as a function of position using ILCBF and GPs.}
	\label{fig:frameworkSchem}
\end{figure}

In this section, feedforward parameters are modeled as a function of position, such that it can compensate for position-dependent effects, hence constituting contribution C1. Furthermore, a data-driven feedforward parameter learning technique is presented, constituting contribution C2. First, GPs are investigated, followed by the application of GPs for feedforward parameters. Third, ILCBF is presented to learn feedforward parameters for a single position. Finally, an example and overview of the framework is presented, which can schematically be seen in Fig.~\ref{fig:frameworkSchem}.
\subsection{Gaussian Processes}
A GP is defined as a collection of random variables $f(\rho)$, indexed by $\rho$, such that the joint distribution of any finite subset of random variables is multivariate Gaussian. A GP is
\begin{equation}
	\label{eq:GPDef}
	f(\rho) \sim \mathcal{GP}\big(m(\rho),k(\rho,\rho^\prime)\big),
\end{equation}
which shows a GP is fully defined by the covariance function $k(\rho,\rho^\prime)$ and the mean function $m(\rho)$,
\begin{equation}
	\label{eq:meanAndCovFunction}
	\begin{aligned}
		k\left(\rho, \rho^\prime\right) &=\mathbb{E}\left[(f(\rho)-m(\rho))\left(f\left(\rho^\prime\right)-m\left(\rho^\prime\right)\right)\right], \\
		m(\rho) &=\mathbb{E}[f(\rho)].
	\end{aligned}
\end{equation}
The mean function $m$ can be interpreted as the mean at any input point and the covariance function $k$ as the similarity between values of $f(\rho)$ on different inputs $\rho$ and $\rho^\prime$.
\subsection{Gaussian Processes for Position-Dependent Feedforward Parameters}
\label{sec:GP}
In this section, feedforward parameters are estimated on test positions $\rho\in P_*$, given feedforward parameters learned on training positions $\rho\in P$ using GPs, leading to contribution C1. The test and training positions are defined as 
\begin{equation}
	\begin{aligned}
		P_* &= \begin{bmatrix}\rho_{1_*} & \rho_{2_*} & \ldots & \rho_{l_*}\end{bmatrix}^\top\in\mathbb{R}^{l_*\times n_o}, \\
		P &= \begin{bmatrix}	\rho_1 & \rho_2 & \ldots & \rho_l\end{bmatrix}^\top\in\mathbb{R}^{l\times n_o},
	\end{aligned}
\end{equation}
with $l_*, \, l \in \mathbb{N}$ the amount of test and training positions, respectively. Observations of the feedforward parameters, i.e. ${\theta}(P)$, are contaminated with noise,
\begin{equation}
	\label{eq:noisyObservations}
	{\theta}(P) = f(P)+\epsilon, \text{ with } \epsilon\sim\mathcal{N}(0,\sigma_\epsilon^2I),
\end{equation}
with the noise variance $\sigma_\epsilon^2$. The continuous function of position $\theta(P_*)$ and the observations $\theta(P)$ are assumed to be random variables and have a joint multivariate Gaussian distribution
\begin{equation}
	\label{eq:jointGaussian}
	\begin{bmatrix}
		{\theta}(P) \\
		{\theta}(P_*)
	\end{bmatrix} \sim \mathcal{N}\left(0,\begin{bmatrix}
	K(P,P)+\sigma_n^2I & K(P,P_*) \\
	K(P_*,P) & K(P_*,P_*)
\end{bmatrix}\right),
\end{equation}
with covariance matrices $K_y:=K(P,P)+\sigma_n^2I \in \mathbb{R}^{l \times l}$, $K_*:=K(P,P_*)=K(P_*,P)^\top \in \mathbb{R}^{l \times l_*}$ and $K_{**}:=K(P_*,P_*) \in \mathbb{R}^{l_* \times l_*}$, which specify similarity between outputs on different positions. The parameter $\sigma_n^2$ is an approximation of the noise variance $\sigma_\epsilon^2$ on the observations in \eqref{eq:noisyObservations}. The parameter $\sigma_n^2$, in addition to the hyperparameters of the covariance function $K$, are approximated based on data by using marginal likelihood optimization \cite[Chapter~5]{Rasmussen2004}. Note that the mean function is assumed to be zero, which is not strictly necessary, see e.g. \cite[Section~2.7]{Rasmussen2004}. The joint distribution in \eqref{eq:jointGaussian} can be conditioned on function observations using Bayes' rule, resulting in the posterior distribution
\begin{equation}
	\label{eq:posteriorDistribuiton}
	{\theta}(P_*)\Big| \big[P,P_*,{\theta}(P)\big] \sim \mathcal{N}\big(\bar{\theta}(P_*),\operatorname{cov}({\theta}(P_*))\big),
\end{equation}
with posterior mean and posterior covariance
\begin{equation}
	\label{eq:postMeanVar}
	\begin{split}
		\bar{\theta}(P_*) := \mathbb{E}(\theta(P_*))&= K_*^\top K_y^{-1}{\theta}(P), \\
		\operatorname{cov}({\theta}(P_*)) &= K_{**}-K_*^\top K_y^{-1}K_*.
	\end{split}
\end{equation} 
In combination with a suitable covariance function, \eqref{eq:postMeanVar} estimates feedforward parameters, given parameter observations.
\subsection{Learning Feedforward Parameters using Basis Functions for a Fixed Position}
\label{sec:ILCBF}
In this section, feedforward parameters are learned for multiple fixed positions to serve as training data for the GP, leading to contribution C2. Here, ILCBF is used to learn the parameters, but the framework can directly be extended to other feedforward parameter tuning approaches. \par
The optimization criterion in ILCBF is specified as \cite{Bolder2014}
\begin{equation}
	\label{eq:perfCritILC}
	\scriptstyle
	V\left({\theta}_{j+1}\right)=\left\|e_{j+1}(\dt)\right\|_{W_{e}}^{2}\!\!+\left\|f_{j+1}(\dt)\right\|_{W_{f}}^{2}\!\!+\left\|f_{j+1}(\dt)-f_{j}(\dt)\right\|_{W_{\Delta f}}^{2},
\end{equation}
with weighting matrices $W_e \succ 0$ and $W_f$, $W_{\Delta f}\succeq 0$ and $\theta_j$ the feedforward parameters in trial $j$. The error $e_{j+1}$ is
\begin{equation}
	\label{eq:errorj+1_2}
	\begin{split}
		e_{j+1}(\dt) &= S(q^{-1})r(\dt)-S(q^{-1})G_0(q^{-1}) f_{j+1}(\dt)\\
		&= e_j(\dt) - S(q^{-1})G_0(q^{-1})\big(f_{j+1}(\dt)-f_j(\dt)\big),
	\end{split}
\end{equation}
where now, $G_0$ is for instance a nominal model of a position-dependent system. The feedforward signal is parameterized in terms of the feedforward parameters $\theta_j$, i.e., $f_j(\dt)=\Psi(q^{-1})r(\dt)\theta_j$. The feedforward parameters are updated as
\begin{equation}
	\label{eq:feedforwardUpdate}
	{\theta}_{j+1}^* = \arg \min_{{\theta}_{j+1}} V\left({\theta}_{j+1}\right).
\end{equation}
Since the feedforward force is chosen linearly in the feedforward parameters, the optimization criterion in \eqref{eq:perfCritILC} becomes quadratic in $\theta_{j+1}$. Hence, an analytic solution to \eqref{eq:feedforwardParameterUpdate} is \cite{Bolder2015}
\begin{equation}
	\label{eq:feedforwardParameterUpdate}
	\begin{split}
		{\theta}_{j+1} &= Le_j+Q{\theta}_j, \\
		L&=R^{-1}\left(\Psi^{\top}G_0^{\top} S^{\top} W_{e}\right), \\
		Q&=R^{-1} \Psi^{\top}\left(G_0^{\top} S^{\top} W_{e} G_0 S+W_{\Delta f}\right) \Psi, \\
		R &= \left(\Psi^{\top}\left(G_0^{\top} S^{\top} W_{e} G_0 S+W_{f}+W_{\Delta f}\right) \Psi\right),
	\end{split}
\end{equation}
where $(q^{-1})$ and $(\dt)$ have been left out for brevity. The parameter update in \eqref{eq:feedforwardParameterUpdate} leads to monotonic convergence of $\|f_j(\dt)\|$, provided matrices $W_e$, $W_f$ and $W_{\Delta f}$ are selected properly \cite{Bolder2014}. Robustness, with respect to model mismatch due to the position-dependent dynamics, can be enforced by increasing $W_f$. Now, \eqref{eq:feedforwardParameterUpdate} can be used in combination with the error in trial $j$ to compute a new set of feedforward parameters $\theta_{j+1}$ for a fixed training position. 
\subsection{Developed Framework and Example}
The combination of GPs and ILCBF is used to model feedforward parameters as a function of position. The framework is schematically shown in Fig.~\ref{fig:frameworkSchem}. First, ILCBF is performed $l$ times on the positions $\rho\in P$, where $P$ is chosen by the user. Second, the feedforward parameters learned in ILCBF, i.e. the training data $\theta(P)$, can be used in a GP regression. Finally, the GP results in the feedforward parameters modeled as a function of the test inputs $P_*$, i.e. $\theta(P_*)$. \par
\begin{example}
	Consider $G_m$ with spatially-distributed mass,
	\begin{equation*}
		\label{eq:exSystem}
		G_{m}(\rho,q^{-1}) = \frac{1}{\bar{m}\left(1-2\left(\frac{1}{2}-\rho\right)^2\right)\frac{(1-q^{-1})^2}{T_s^2}} \quad \forall \rho \in [0,1],
	\end{equation*}
	with $\bar{m}=1$ kg a nominal mass. The feedforward controller is designed using acceleration feedforward, i.e.,
	\begin{equation*}
		f_j(k) = \Psi(q^{-1})r(\dt)\theta_j = \ddot{r}(\dt)\theta_j.
	\end{equation*}
	The feedforward parameters are learned on 4 positions $\rho\in P$, with $P = \begin{bmatrix}0.05 & 0.35 & 0.65 & 0.95\end{bmatrix}$, using ILCBF. The feedforward parameters have bias with respect to the true system values due to measurement noise, which is typically observed in ILCBF \cite{Boeren2015}. The framework is applied using a squared exponential covariance function, see e.g. \cite[Section~4.2]{Rasmussen2004}. Fig.~\ref{fig:GPregressionExample} shows how the framework models a position-dependent feedforward parameter, resulting in an accurate model.
	\begin{figure}[!t]
		\centering
		\includegraphics{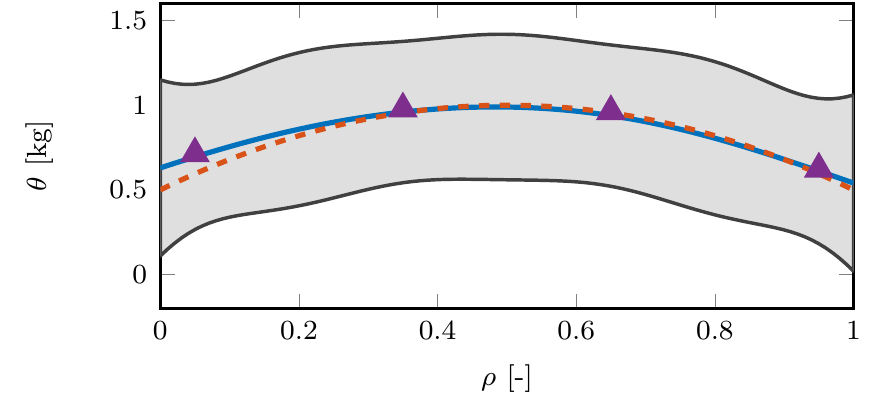}
		\vspace{-8mm}
		\caption{Example GP regression created using the framework of a system with a spatially-distributed mass. The training data (\filledTriangle{purple}) is used for the posterior mean (\protect\blueline) from \eqref{eq:postMeanVar} and the uncertainty bound $\bar{\theta}(P_*)\pm 2 \sigma$ (\protect\fillGP). An accurate representation of the true mass (\protect\redline) is achieved with the GP regression.}
		\label{fig:GPregressionExample}
	\end{figure}
\end{example}

%% file: ExperimentalStudy.tex
\section{Experimental Case Study}
\label{sec:experiments}
In this section, position-dependent feedforward with GPs is applied to a benchmark experimental setup, hence constituting contribution C3. First, the example setup is presented, followed by the application and results of the framework. Finally, a discussion is presented regarding the outcome of the experiments.
\subsection{Experimental Setup}
Position-dependent effects hamper the performance of semiconductor back end equipment, such as the commercial wire bonder seen in Fig.~\ref{fig:AB383}. Position-dependent dynamics can be caused due to several reasons, e.g. position-dependent actuators, flexible dynamics and changing configuration. The wire bonder consists of two inputs $u_1$ and $u_2$ and two outputs $y_1$ and $y_2$. The feedforward parameters are modeled as a function of the initial position of the machine for $y_1$ and $y_2$. The feedforward force $f$ is parameterized using
\begin{equation}
	\label{eq:bfmatrix}
	\Psi(q^{-1})r(\dt) = \begin{bmatrix}
		\scriptstyle\dot{r}_1(\dt) & \scriptstyle\ddot{r}_1(\dt) & \scriptstyle0 & \scriptstyle0 & \scriptstyle0 \\
		\scriptstyle0 &\scriptstyle 0 &\scriptstyle \dot{r}_2(\dt) &\scriptstyle \ddot{r}_2(\dt) &\scriptstyle \psi_2(\dt)
	\end{bmatrix},
\end{equation}
with $\psi_2(\dt)$ a non-linear basis function. The references $r_1$ and $r_2$ are polynomial trajectories, see \cite{Lambrechts2005}. The feedforward parameters are defined as
\begin{equation}
	\label{eq:FFParameters}
{\theta} = \begin{bmatrix}
		\theta_1 & \theta_2 & \theta_3 & \theta_4 & \theta_5
	\end{bmatrix}^\top,
\end{equation}
which are used, in combination with the basis function matrix in \eqref{eq:bfmatrix}, to calculate the feedforward force $f$.
\subsection{Experimental Results}
\begin{figure}[tbp]
	\centering
	\includegraphics{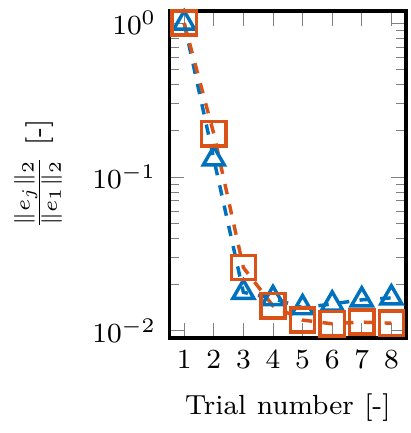}
	\includegraphics{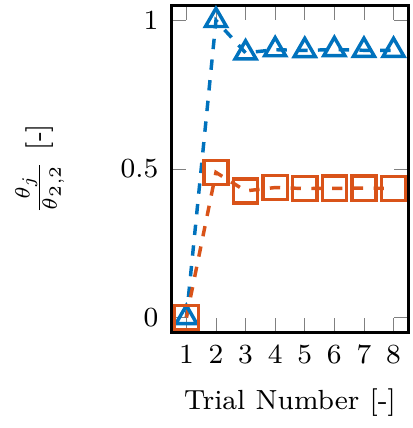}
	\vspace{-4mm}
	\caption{Left: Normalized error 2-norm for the first 8 trials of ILCBF in the center position for $y_1$ (\triangleDash{blue}) and $y_2$ (\squareDash{red}). Right: Acceleration feedforward parameters $\theta_2$ (\triangleDash{blue}) and $\theta_4$ (\squareDash{red}) for the first 8 trials of ILCBF in the center position, normalized with respect to parameter $\theta_2$.}
	\label{fig:ILCBFCenter}
\end{figure}

\begin{figure}[tbp]
	\centering
	\hspace{-7mm}
	\includegraphics{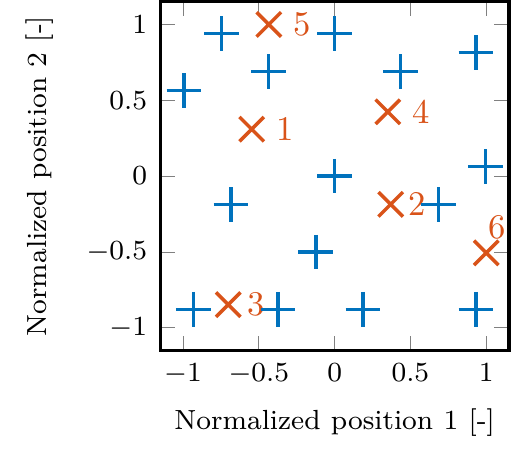}
	\vspace{-4mm}
	\caption{ILCBF will be performed on the arbitrarily chosen training positions (\plusMarker{blue}) and the learned feedforward parameters on these positions will be used as training data for the GPs. The framework will be tested and compared on the test positions (\crossMarker{red}), consisting of interpolation and extrapolation on the data.}
	\label{fig:trainingAndTestPositions}
\end{figure}

\begin{figure}[tbp]
	\centering
	\includegraphics{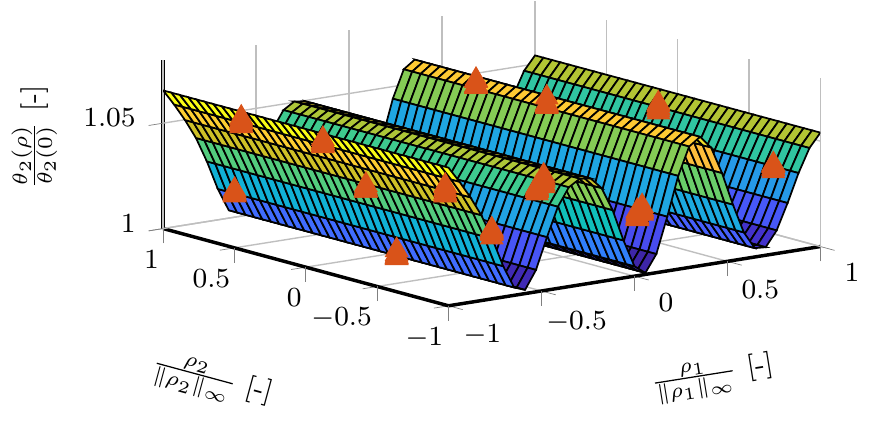}
	\vspace{-8mm}
	\caption{Normalized GP regression of the acceleration feedforward parameter $\theta_2$, representing the estimated mass of the axis, made using the training data (\filledTriangle{red}). The feedforward parameter has a clear dependency on $\rho_1$.}
	\label{fig:GP2}
\end{figure}

\begin{figure}[tbp]
	\centering
	\includegraphics{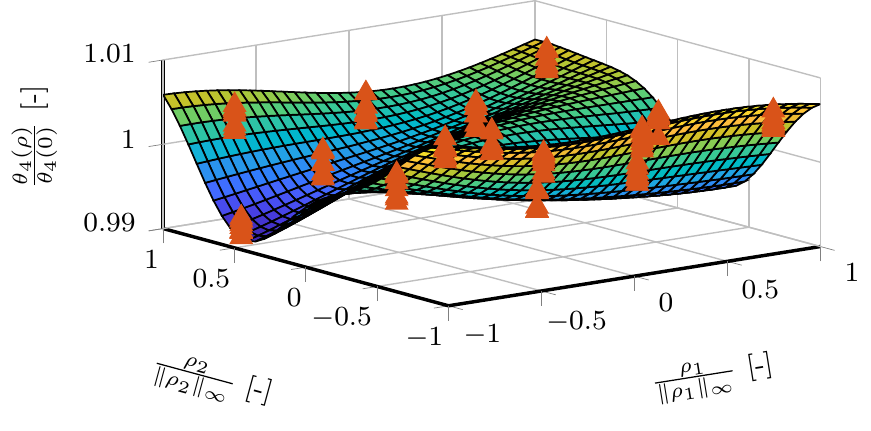}
	\vspace{-8mm}
	\caption{Normalized GP regression of the acceleration feedforward parameter $\theta_4$ as a function of position, representing the estimated mass of the axis, made using the training data (\filledTriangle{red}). Position-dependency is seen for $\rho_1$ and $\rho_2$.}
	\label{fig:GP4}
\end{figure}

The framework is applied to the wire bonder seen in Fig.~\ref{fig:AB383}. First, feedforward parameters are learned on several positions with ILCBF using 20 trials, as described in Section~\ref{sec:ILCBF}. The error 2-norm and parameters of the first 8 trials in the center position can be seen in Fig.~\ref{fig:ILCBFCenter}, which shows that ILCBF has converged. 
Second, each feedforward parameter in \eqref{eq:FFParameters} is modeled as a separate GP, as seen in Section~\ref{sec:GP}, by using the feedforward parameters learned in trials 13 up to 20 on the training positions in Fig.~\ref{fig:trainingAndTestPositions}. As many converged feedforward parameters should be used in the algorithm, such that the result is unbiased and the variance of the GP is as low as possible. A squared exponential covariance function is used.
For visualization purposes, the positions $P_*$ are chosen as a fine grid covering the operating range of the machine. In Fig.~\ref{fig:GP2} and Fig.~\ref{fig:GP4} the GP regressions of the acceleration feedforward parameters can be seen and show a significant difference in the parameters for the operating range of the machine. \par
\begin{figure}[tbp]
	\centering
	\includegraphics{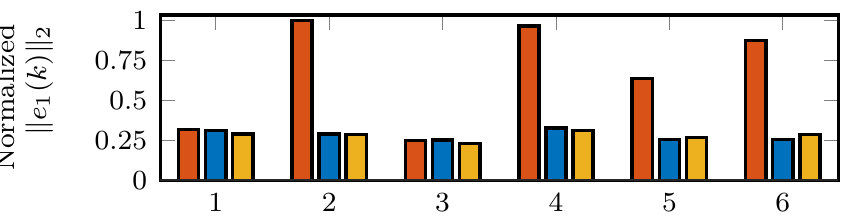}
	\includegraphics{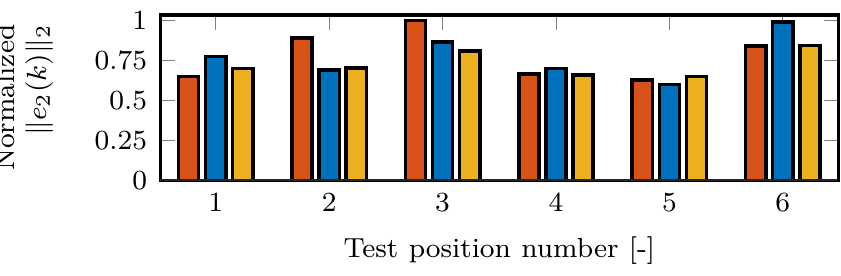}
	\vspace{-4mm}
	\caption{Top: $\|e_1(k)\|_2$ for center (\protect\fillred), GP (\protect\fillblue) and local ILCBF (\protect\fillyel) feedforward. Both GP and ILCBF achieve constant performance, while center has significantly higher error 2-norm for several positions.	Bottom: Error 2-norm $\|e_2(k)\|_2$ for center (\protect\fillred), GP (\protect\fillblue) and local ILCBF (\protect\fillyel) feedforward. The performance for $y_2$ is similar for center, GP and ILCBF feedforward.}
	\label{fig:error2Norm}
\end{figure}

\begin{figure}[tbp]
	\centering
	\includegraphics{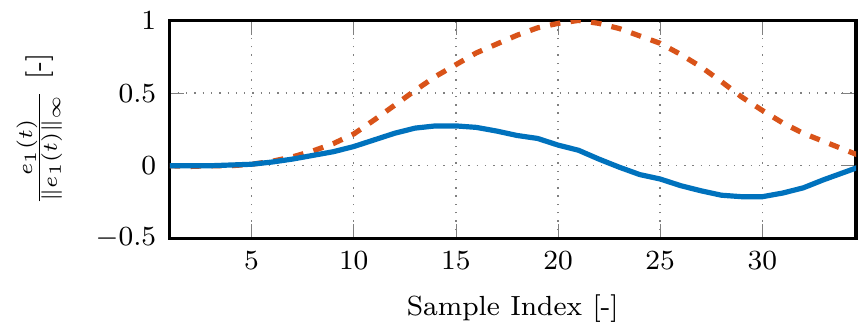}
	\vspace{-4mm}
	\caption{Time-domain error $e_1(k)$ for test position 2, for both center (\protect\redline) and GP (\protect\blueline) feedforward. The maximum error for GP feedforward is roughly 3 times smaller than center feedforward.}
	\label{fig:error2}
\end{figure}

To evaluate the performance of the framework, several test positions other than the training positions have been chosen, seen in Fig.~\ref{fig:trainingAndTestPositions}. The GP regressions made with the feedforward parameters learned on the training positions are used to estimate feedforward parameters on the test positions using \eqref{eq:postMeanVar}. On the test positions, three methods of choosing feedforward parameters have been performed:
\begin{enumerate}
	\item[\textbf{M1}] center, where the feedforward parameters determined in the center position of the machine are used, 
	\item[\textbf{M2}] GP, where the feedforward parameters are estimated using the GP at the given test position,
	\item[\textbf{M3}] local ILCBF, where the feedforward parameters are determined at the test position using ILCBF.
\end{enumerate}
Here, local ILCBF is performed to serve as a reference frame to compare the first two methods with. The error 2-norm for the test positions for all methods can be seen in Fig.~\ref{fig:error2Norm}. The error 2-norm of $y_1$ in Fig.~\ref{fig:error2Norm} shows that a significant performance increase can be achieved when using feedforward parameters modeled as a function of position by a GP. This is additionally supported by looking at the time domain error of $y_1$ for test position 2 in Fig.~\ref{fig:error2}. Furthermore, the GP feedforward achieves an error 2-norm similar to that of ILCBF, indicating optimal performance in terms of \eqref{eq:perfCritILC} with the specified basis function and requiring no learning at the test position, in contrast to local ILCBF. The error 2-norm of $y_2$ shows the performance difference for the two methods and ILCBF is only marginal, showing that the $y_2$ direction does not have much position-dependency. 
\subsection{Discussion of Experimental Case Study}
The experimental setup has position-dependent effects, which can be concluded by looking at the feedforward parameters in Fig.~\ref{fig:GP2} and Fig.~\ref{fig:GP4}. For the acceleration feedforward parameter $\theta_2$ in Fig.~\ref{fig:GP2}, significant position-dependent effect is seen. The unusual behavior of the parameter might be caused due to the periodic magnetic flux density in linear actuators \cite{Gieras2012}. Indeed, when measuring the distance between the peaks in Fig.~\ref{fig:GP2}, a distance roughly equal to the magnet pitch of the linear actuator is observed. The position-dependent effect is directly seen in the performance difference for $y_1$ in Fig.~\ref{fig:error2Norm}. \par 
For test positions 1 and 3, the performance for center and GP feedforward are equal for the $y_1$ direction. This can directly be explained by looking at Fig.~\ref{fig:trainingAndTestPositions} and Fig.~\ref{fig:GP2}, that shows test positions 1 and 3 are located roughly one magnet pitch away from the center position, resulting in optimal feedforward parameters equal to the center position. \par 
Compared with the acceleration feedforward parameter $\theta_2$, the acceleration feedforward parameter $\theta_4$ in Fig.~\ref{fig:GP4} has considerably less position-dependency. The unusual behavior is also not observed for $\theta_4$, indicating that the magnetic flux density is not affecting the feedforward parameter as much. When taking a look at the performance for $y_2$, the difference is only marginal between the feedforward methods. Therefore, it can be concluded that for the considered experimental setup, position-dependent effects in the actuator heavily outweigh position-dependent effects in the mechanics or sensing.